\documentclass{rnaastex}
\usepackage{url}

\begin{document}

\title{RV-detected \textit{Kepler}-multi Analogs Exhibit Intra-system Mass Uniformity}

\correspondingauthor{Songhu Wang}
\email{song-hu.wang@yale.edu}

\author{Songhu Wang}
\affiliation{Department of Astronomy, Yale University, New Haven, CT 06511 }
\affiliation{51 Pegasi b Fellow}


\begin{abstract}

Recent work has shown that the planets in the $Kepler$ Mission's population of multi-transiting systems show surprising uniformity in both mass and radius. In this brief note, I show that this intra-system mass uniformity extends to multiple-planet systems detected with the Doppler velocity technique, thereby avoiding possible biases associated with masses determined by transit timing. I also show that intra-system mass uniformity breaks down when a system contains one or more giant planets.

\end{abstract}

\section{} 

The \textit{Kepler} Mission's most startling discovery was the realization that $\sim50\%$ of Solar-type stars are accompanied by short-period planets with $M_{\rm P}\lesssim 30\,M_{\oplus}$ \citep{Lissauer2014}. These worlds, which often appear in multi-transiting configurations, are quite ordered, displaying low mutual inclinations, low eccentricities, and low spin-orbital misalignments \citep{Winn2015}. In aggregate, however, they generate a highly scattered mass-radius relation. \citep{Weiss2014}.

A recent article by \citet{Weiss2017} emphasized that the \textit{Kepler} multi-transiting systems exhibit a surprising degree of intra-system uniformity in the planetary radii. \citet{Millholland2017} extended this statistically significant ``peas-in-a-pod'' trend to intra-system masses by drawing on the selection of \textit{Kepler} planets for which transit timing variations (TTVs) can determine masses. Such ``TTV'' planets, however, are usually near mean-motion resonance \citep[e.g.,][]{Lithwick2012}, leading naturally to concern over whether a finding that holds for a potentially special class of planets can be generalized. Indeed, the discrepancy between RV- and TTV-determined masses suggests there may be potential physical differences that are correlated with proximity to resonance \citep{Steffen2016,Mills2017}.

The prospects for usefully extending \citet{Millholland2017}'s set of mass determinations are rather dim. A significant number of additional masses would require many precise Doppler velocity (RV) measurements, which are impractical for the faint \textit{Kepler} targets. Several groups, however, have consistently achieved long-term $1-3\,{\rm m\,s^{-1}}$ Doppler precision on bright, stable stars \citep{Mayor2009, Vogt2010, Howard2011}. The extant catalog of Doppler-discovered multi-planet systems thus provides us with an independent sample of \textit{Kepler}-multi analogs to assess the degree of intra-system mass uniformity among planets that are far from mean-motion resonance.

We draw from \url{https://exoplanetarchive.ipac.caltech.edu/}, to establish a sample of the 29 known systems with at least three Doppler-detected planets. Two-planet systems are omitted because they often display large period ratios, and may be not co-planar. 

\begin{figure}[h!]
\begin{center}
\includegraphics[scale=0.5,angle=0]{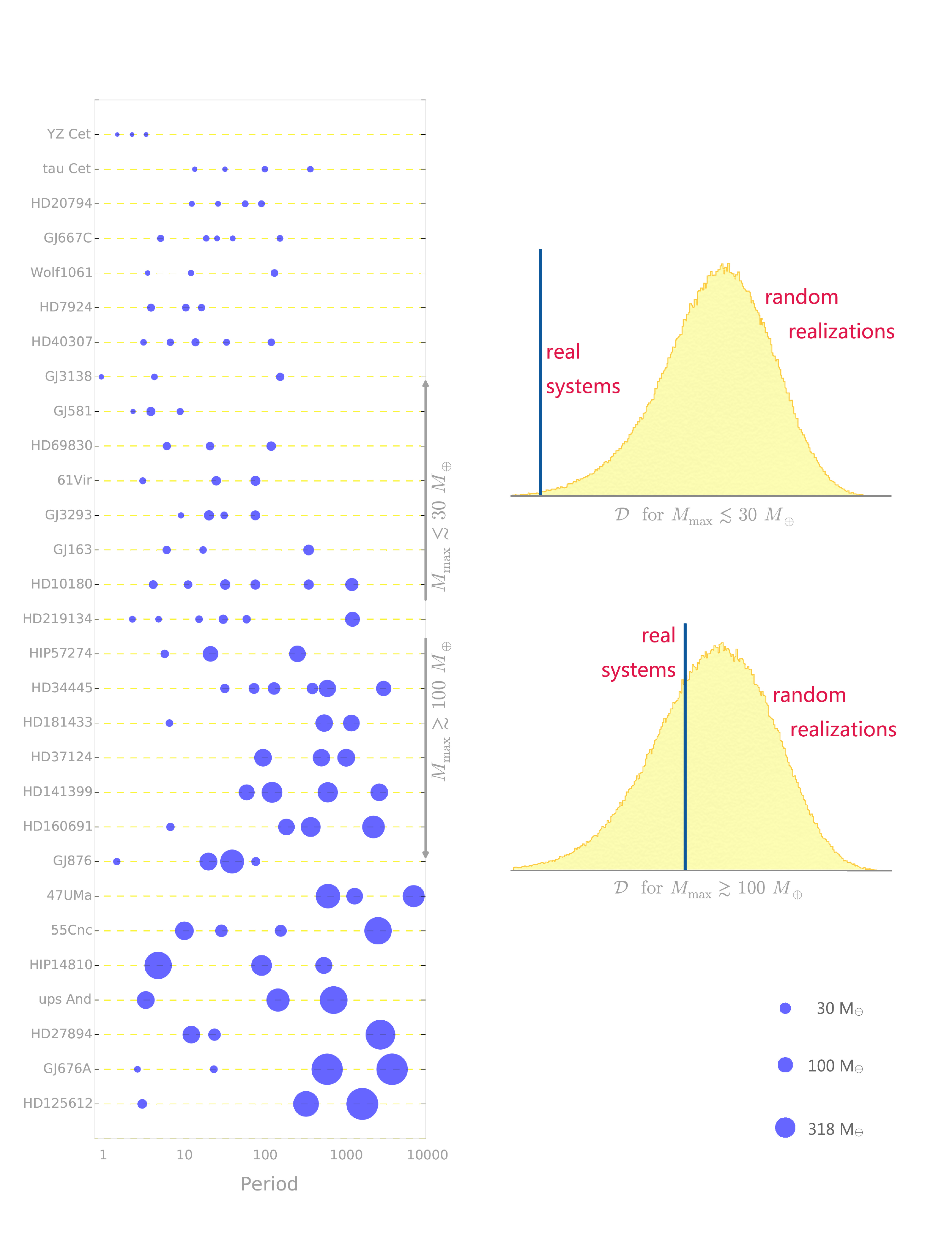}
\caption{\textit{Left}: Architectures of RV-detected planetary systems with at least 3 planets. The aggregate is ordered by the maximum $M\sin(i)$ in the system. The systems are divided into two groups. In the first group, $M_{\rm max} \lesssim 30\,{M_\oplus}$. These systems resemble the population in the \textit{Kepler} census, and exhibit significant intra-system mass uniformity. The second group has systems with  $M_{\rm max} \gtrsim 100\,{M_\oplus}$. This group shows considerable intra-system dispersion.
\textit{Right}: Comparisons of the dispersion metric $\mathcal{D}$ between the real systems (vertical blue lines) and the control populations of 50,000 realizations of shuffled systems (yellow histograms).
 \label{fig:1}}
\end{center}
\end{figure}

The set of systems, ordered by $M\sin(i)_{\rm max}$, is shown in Figure~\ref{fig:1}, and divides naturally into two groups. In the first, the maximum planetary mass is below the core rapid gas-accretion threshold of $\sim 30\,{M_\oplus}$. These systems resemble the dominant planetary population in the \textit{Kepler} census, with a prototypical example being HD 40307 \citep{Mayor2009}. The second class contains gas giants with planetary masses greater than $\sim 100\,{M_\oplus}$. A representative member of this group is 55 Cnc \citep{Fischer2008}.

To quantify whether planets in a given system are preferentially correlated in mass, we define the intra-system mass dispersion $\mathcal{D}$ as

\begin{equation}
\label{radius dispersion metric}
\mathcal{D} = \sum_{i=1}^{N_{\mathrm{sys}}}  
\sqrt{\frac{\sum_{j=1}^{N_{\mathrm{pl}}}    (M_{j}-\overline{M})^{2}}{N_{\mathrm{pl}}-1}}\, ,
\end{equation}
where $\overline{M}=(1/N){\sum_{j=1}^{N_{\mathrm{pl}}}M_{j}}$. Figure~\ref{fig:1} shows that $\mathcal{D}$, as exhibited by the small-planet systems (vertical blue line) is a $3.2\sigma$ outlier in comparison to a Monte-Carlo control sample (yellow histogram) in which the number of planets in each system and the total number of systems were conserved, but the planets were randomly shuffled 50,000 times. By contrast, the frequent co-existence of small planets and gas giants in the second class produces significant intra-system mass dispersions. For this group, the real systems show no significant departure from random shuffling. 

The intra-system mass uniformity that characterizes the $Kepler$ TTV sample thus extends to RV-detected $Kepler$-multi analogs, which contain sets of small close-in planets, but which avoid the TTV mass determination bias toward near-resonances. This result, together with those of \citet{Weiss2017} and \citet{Millholland2017}, implies that planet formation within a given system is a coordinated process. One notes a similarity to the Solar System's Jovian satellites the small scale ($\sim 0.01\,$AU) and to the galactic conformity on the large scale ($\sim1\,$Mpc, \citealt{Weinmann2006}). The oft-remarked diversity among exoplanets is perhaps better expressed as a diversity among exoplanetary \textit{systems}.

\acknowledgements

I thank the Heising-Simons Foundation for their generous support. I would also like to thank Greg Laughlin for inspirational conversations and editing. Additionally, I am grateful to Yutong Wu for improving the quality of the Figure, as well as to Sarah Millholland, Song Huang, Zhongyi Man, and Frank van den Bosch for useful discussion. 

\end{document}